\documentclass[twocolumn]{article} 

\newcommand{\etal}{\emph{et al.}}
\newcommand{\eg}{\emph{e.g. }}
\newcommand{\ie}{\emph{i.e. }}

\newcommand{\SC}{{\sf SC}}
\newcommand{\PSS}{{\sf SS}}
\newcommand{\RS}{{\sf RS}}
\newcommand{\RC}{{\sf RC}}

\newcommand{\SD}{{\sf SD}}
\newcommand{\RD}{{\sf RD}}

\newcommand{\website}{\url{http://www2.kinneret.ac.il/mjmay/wsemail/}}
\newcommand{\code}{\url{https://github.com/lux-k/wsemail}} 

\usepackage{ifthen}
\newboolean{tr}
\newcommand{\tr}{\ifthenelse{\boolean{tr}}}

\usepackage{graphicx, url, color,fullpage,booktabs}
\usepackage[numbers]{natbib}

\title{WSEmail: A Retrospective on a System for Secure Internet Messaging Based on Web Services}

\author{\and Michael J. May\\Kinneret Academic College\\mjmay@kinneret.ac.il 
        \and Kevin D. Lux\\University of Pennsylvania, Rowan University\\
              kevin@kevinlux.info               
            \and Carl A. Gunter\\University of Illinois Urbana-Champaign\\cgunter@illinois.edu}

\date{}

\begin{document}

\maketitle

\begin{abstract}
Web services offer an opportunity to redesign a variety of older systems to exploit the advantages of a flexible, extensible, secure set of standards. In this work we revisit WSEmail, a system proposed over ten years ago to improve email by redesigning it as a family of web services.  WSEmail offers an alternative vision of how instant messaging and email services could have evolved, offering security, extensibility, and openness in a distributed environment instead of the hardened walled gardens that today's rich messaging systems have become.  WSEmail's architecture, especially its automatic plug-in download feature allows for rich extensions without changing the base protocol or libraries.  We demonstrate WSEmail's flexibility using three business use cases: secure channel instant messaging, business workflows with routed forms, and on-demand attachments.  Since increased flexibility often mitigates against security and performance, we designed WSEmail with security in mind and formally proved the security of one of its core protocols (on-demand attachments) using the TulaFale and ProVerif automated proof tools. We provide performance measurements for WSEmail functions in a prototype we implemented using .NET. Our experiments show a latency of about a quarter of a second per transaction under load.
\end{abstract}

\paragraph{Keywords}{Internet electronic mail, web services, WSEmail, security, performance, rich messaging, on-demand attachments, email workflow}

\section{Introduction}

Web services are a mature technology nearing their twentieth birthday.  They have created the foundation for highly interoperable distributed systems to communicate over the Internet using standardized protocols (\eg SOAP, JSON) and security mechanisms (\eg OAuth (IETF RFC 6749), XMLDSIG~\cite{xmldsig}). Legacy systems and protocols must be reevaluated to see how they can benefit from modern architectures, standards, and tools. As a case study of such an analysis and redesign, we present an expanded study of WSEmail~\cite{LuxMBG05}, electronic mail redesigned as a family of web services we first implemented and presented in 2005.

Such a return is warranted due to a consideration of how internet messaging technologies have evolved in the past decade and a half.  When we first implemented WSEmail, email and instant messaging (IM) services were strictly disjoint.  Instant messaging solutions (\eg AOL Instant Messenger, ICQ) were server centric, offered little to no security, had weak authentication, and worked only when both sides were online.  Email services were more mature with endpoint security and authentication options, but in order to support uniformity across a vast installed base, the resulting system had shortcomings in the areas of flexibility, security, and integration with other messaging systems. For instance, problems of remote authentication and extensibility plagued attempts to reduce spam, while poor integration with browsers and operating systems made it a vector for the propagation of viruses, malware, and ransomware.  In the intervening years, IM and email have evolved separately.

Email has become hardened with the standardization of spam blocking (\eg real time black hole lists, policy block lists), DomainKeys Identified Mail (DKIM) (IETF RFC 6376), Domain Name System Security Extensions (DNSSEC) (IETF RFC 4033-5), and encryption by default between Mail Transfer Agents (MTAs). Together, they made email more secure in transit and reduced the quantity of received spam.  Push notifications changed the speed at which users see email, but the fundamental message format \tr{(7-bit ASCII with MIME)} is unchanged.  Importantly, it has remained an open system with distributed management and no central point of control.

In parallel, IM underwent fundamental changes as the new generation of tools (\eg Facebook Messenger, WhatsApp, Skype, Slack, WeChat) introduced stronger authentication, end-to-end security, and rich communications features such as bots, mini-applications, and video chat.  IM apps cache messages sent or received while offline and can show proof of delivery.  In contrast to email services, IM networks have become ``closed gardens'' with little to no interoperability.  With few exceptions, access is available only via dedicated clients via a central point of control.  Support for offline sending and receiving blur the boundary between IM and email, but IM security protocols (\eg Signal Protocol~\cite{Signal16}) are centralized, preventing distributed management and customizations.  No integration with email is possible.

It didn't have to be this way.

We built WSEmail to replace legacy protocols with protocols based on SOAP, WSDL, XMLDSIG and other XML-based formats. The protocols are proven technologically (they have been standardized for over 15 years) and they are inherently extensible, give stronger guarantees for message authentication, and are amenable to formal modeling and proof.

WSEmail is designed to perform the functions of ordinary email but also enable additional security functions and flexibility. The primary strategy is to import these virtues from the standards and development platforms for web services. Our exploration of WSEmail is based on a prototype architecture and implementation. WSEmail messages are SOAP messages that use web service security features to support integrity, authentication, and access control for both end-to-end and hop-by-hop message transmissions. The WSEmail platform supports the dynamic updating of messaging protocols on both client Mail User Agents (MUAs) and server MTAs to enable custom communications. This flexibility supports the introduction of new security protocols, richer message routing (such as routing based on the semantics of a message), and close integration with diverse forms of communication such as IM.

The benefits of flexibility can be validated by showing diverse applications. We show the flexibility of WSEmail in three applications we implemented based on its framework: secure instant messaging, secure business workflow messaging, and on-demand attachments, in which email with an attachment leaves the attachment on the sender's server rather than placing it on the servers of the recipients.  We achieve all of this while avoiding becoming a ``closed garden'' by explicitly considering extensibility and on-demand download of client extensions.  This allows endpoint servers to design their own rich messaging extensions and deploy them locally while maintaining interoperability with others.

Flexibility, however, often has a high cost for security and performance.  We therefore develop techniques to measure and mitigate these costs for WSEmail. WSEmail's first contribution was a case study of a formal analysis of on-demand attachments. The challenge was to design the security for the attachment based on emer\-ging federated identity systems. Due to space restrictions, the proof is not reproduced here, but can be found in Lux~\etal~\cite{LuxMBG05} and in an online appendix (see below).  In this work, we first detail WSEmail's architecture and three of its applications. They show the flexibility we can achieve using our architecture while providing strong security guarantees.  Second, we carry out a set of experiments to determine the efficiency of our base system, including its security operations.  Since email systems need to also have good performance on older hardware, we show our experiments on a testbed built from older hardware and systems.  Both studies demonstrate promise for the security and performance of WSEmail.

WSEmail shows an alternative evolutionary path email and IM could have taken, keeping advantages from both and giving a path for future growth and improvement.

The paper is organized as follows. First, we sketch WSEmail's architecture, focusing on its security assum\-ptions and continue with a discussion about how plug-ins function. Section~\ref{sec:interfaces} discusses interface details of the WSEmail base architecture, including detailed descriptions of how messages are sent and received.  Section~\ref{sec:applications} discusses applications we have explored with WSEmail, including IM, semantic based routing for business workflows, and on-demand attachments.  Section~\ref{sec:experiments} presents our implementation and its performance. Section~\ref{sec:related} has related work and compares WSEmail to similar messaging systems. Section~\ref{sec:conclusion} concludes. Interested readers can find more information in the online appendix at \website.

\section{Base Architecture}\label{sec:base}

\begin{figure}
  \begin{center}
    \includegraphics[width=\columnwidth]{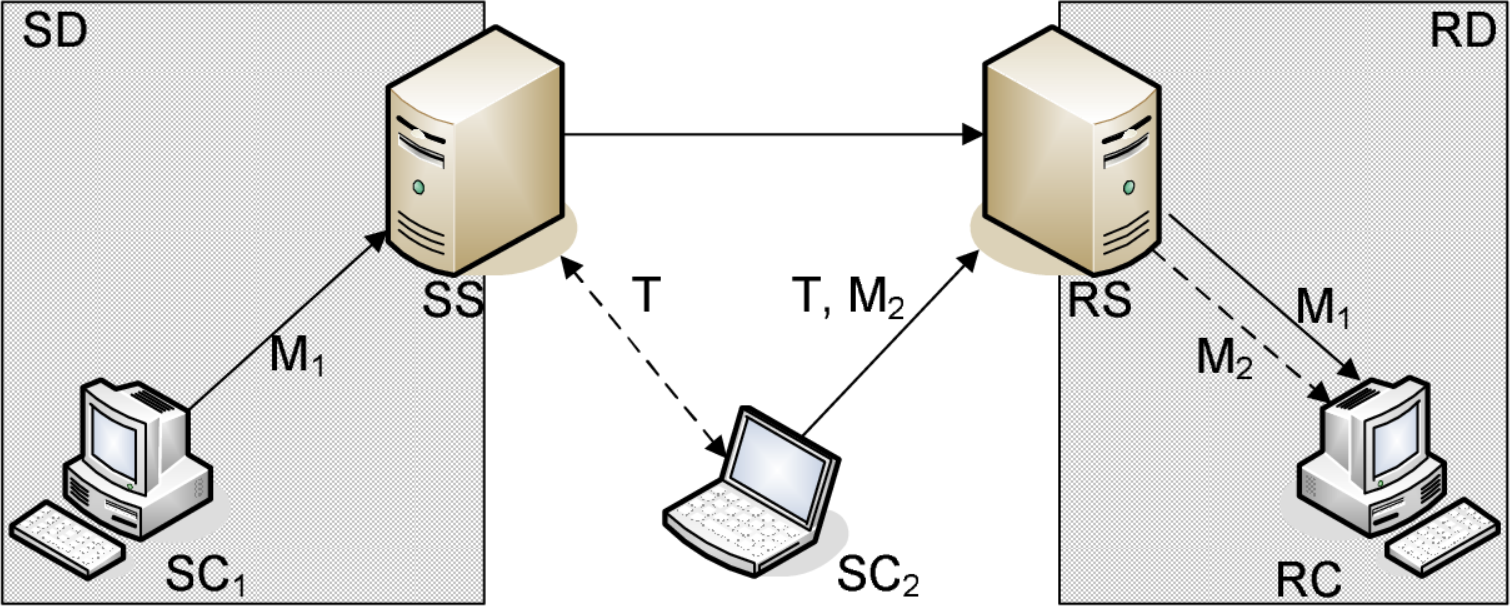}
    \caption{Messaging architecture}
    \label{fig:arch}
  \end{center}
\end{figure}

The base protocols for WSEmail are illustrated in Figure~\ref{fig:arch}. In the common case, similar to SMTP, an MUA Sender Client $\SC_1$ makes a call on its MTA Sender Server $\PSS$ to send a message $M_1$. This and other calls are SOAP calls over TCP; the message $M_1$ is in the body of the SOAP message and the SOAP header contains information like the type of call and security parameters. The message is structured as a collection of XML elements, including, for instance, a subject header. A sample trace of WSEmail messages can be found in the online appendix. After receiving the call from $\SC_1$, the server $\PSS$ makes a call on the Receiver Server $\RS$ to deliver the mail from the Sender Domain $\SD$ to the Receiver Domain $\RD$. The Receiver Client $\RC$ makes calls to $\RS$ to inquire about new messages or download message bodies. In particular, $\RC$ makes a call to $\RS$ to obtain message headers and then can request $M_1$.

\begin{figure}
\centering
\includegraphics[width=\columnwidth]{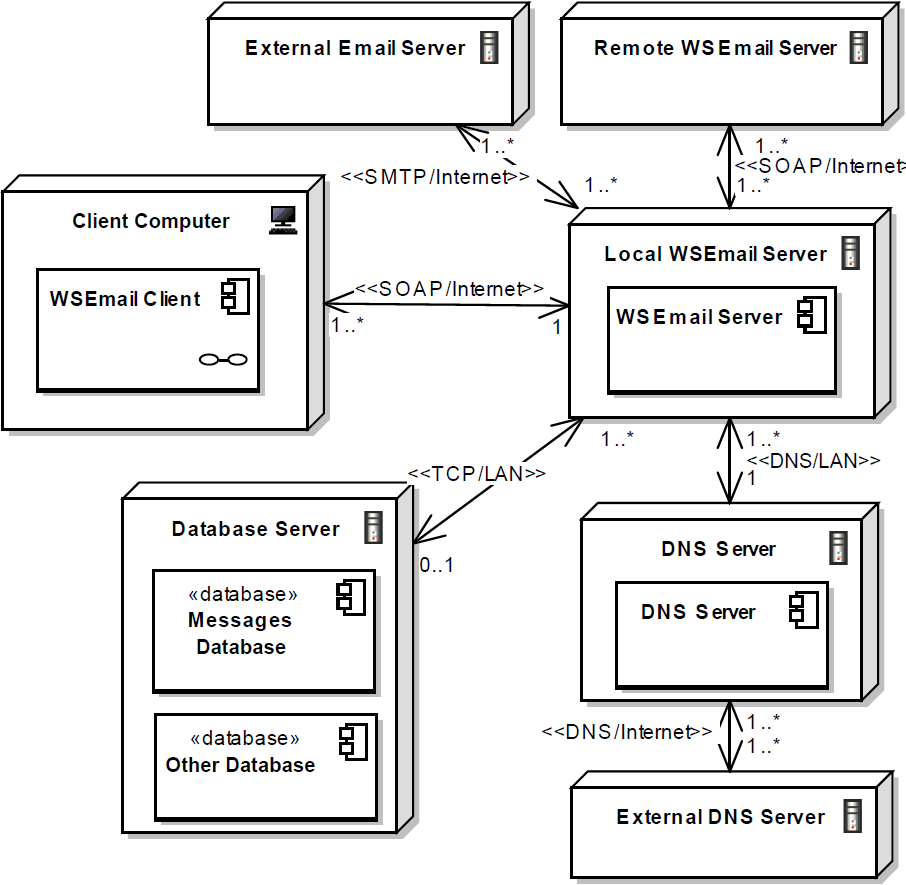}
\caption{WSEmail Deployment Diagram}\label{fig:deployment}
\end{figure}

The deployment diagram for WSEmail's client and server and the associated nodes they use is shown in Figure~\ref{fig:deployment}.

\paragraph{Security Architecture}
Our design is based on a three-tier authentication system combined with an extensible system of federated identities. The first tier provides user (MUA) authentication based on passwords, public keys, or federated identity tokens. The second tier provides server (MTA) authentication based on public keys with certificates similar to those used for Transport Layer Security (TLS). The third tier uses root certificates similar to the ones in browsers. Overall, this addresses interdomain authentication in a practical way at the cost of full end-to-end confidentiality. Confidentiality is preserved between hops by TLS or another tunnel protocol. In a basic instance, the message from $\SC_1$ to $\RC$ will be given an XMLDSIG signature by $\PSS$ that is checked by both $\RS$ and $\RC$.

The novel aspects of WSEmail's architecture are in the integration and flexibility of the MUA authentication and the ability of both MUAs and MTAs to add new security functions dynamically. To illustrate a variation in the base protocol, consider our design for IM. Referring again to Figure~\ref{fig:arch}, an instant message $M_2$ is dispatched from a client $\SC_2$ to $\RC$ while $\SC_2$ is outside its home domain $\SD$. In this case $\SC_2$ contacts $\PSS$ to obtain a security token T that will be recognized by $\RS$. Once it is obtained, $\SC_2$ sends $M_2$ authenticated with the credential to $\RS$ and indicates (in a SOAP header) that it should be treated as an instant message by $\RS$ and $\RC$. Instant messages are posted directly to the client, with the client now viewed as a server that accepts the instant message call. $\RS$ and $\RC$ are able to apply access control for this function based on the security token from $\SC_2$. The token is recognized because of a prior arrangement between $\PSS$ and $\RS$.

\begin{figure*}[pht]
\centering
\includegraphics[width=\textwidth]{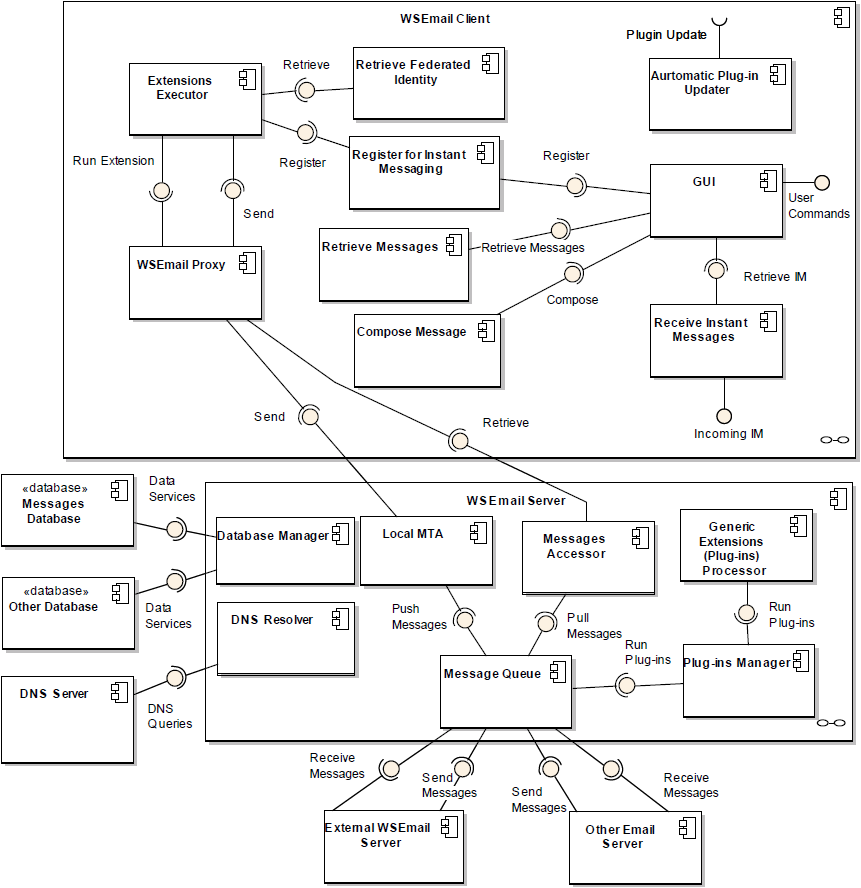}
\caption{Component diagram for WSEmail client, server, and associated services.  The plugins are elaborated on in Figure~\ref{fig:server-plugins-component}}
\label{fig:component-diagram}
\end{figure*}

\paragraph{Plug-ins}
The WSEmail MUA and MTA are based on a plug-in architecture capable of dynamic extensions. Security for such extensions is provided though a policy for trusted sources and the enforcement mechanisms provided by web services. On-demand attachments are an example of such a plug-in, as are a variety of kinds of attachments with special semantics. A party that sends a message with such an attachment automatically includes information for the receiver about where to obtain the software necessary to process the attachment. The client provides hooks for plug-ins to access security tokens, after first performing an access control check on the plug-in. 

\paragraph{Composite Architecture}
The composite architecture for the MUA (client) and server (MTA) is shown in Figure~\ref{fig:component-diagram}.  The server components shown perform the following tasks.

\begin{description}
\item[Database Manager] Provides access to the database for message storage and retrieval.
\item[DNS Resolver] Provides an interface to the DNS system.
\item[Plug-ins Manager] Allows for increased functionality without changing the base server code.  We detail how they work in Section~\ref{sec:server-plugins} below.
\item[Message Queue] Queues messages for local and remote delivery.
\item[Local MTA] Serializes messages to a local data store (\eg saving to a spool file).
\end{description}


\section{Interfaces and Plug-ins}\label{sec:interfaces}

To explain WSEmail's architecture, we outline its code and plug-in architecture and interfaces. The interfaces describe how much access the application has to the mail server or client software, in addition to where the plug-in is activated for message delivery or reading.  Plug-ins are used on both the server and the client.  \tr{Their interactions and communication paths are the source of WSEmail's flexibility and extensibility.}{}

\subsection{Server-side Plug-ins}\label{sec:server-plugins}

Server plug-ins are code libraries that conform to interfaces known to the server. When the server is initialized, it goes through a list of plug-ins to load from a configuration file. For each plug-in, an object class is listed along with the name of a library from which it can be loaded. The server looks for the library and tries to instantiate and load the object. If successful, the server requests further configuration data from the plug-in and uses it to put the plug-in in the appropriate execution queue. The process of loading or unloading plug-ins is dynamic; they can be loaded or unloaded at any time during execution. The execution queues can also be reprioritized or disabled at run time.

\begin{figure}[ht]
\centering
\includegraphics[width=\columnwidth]{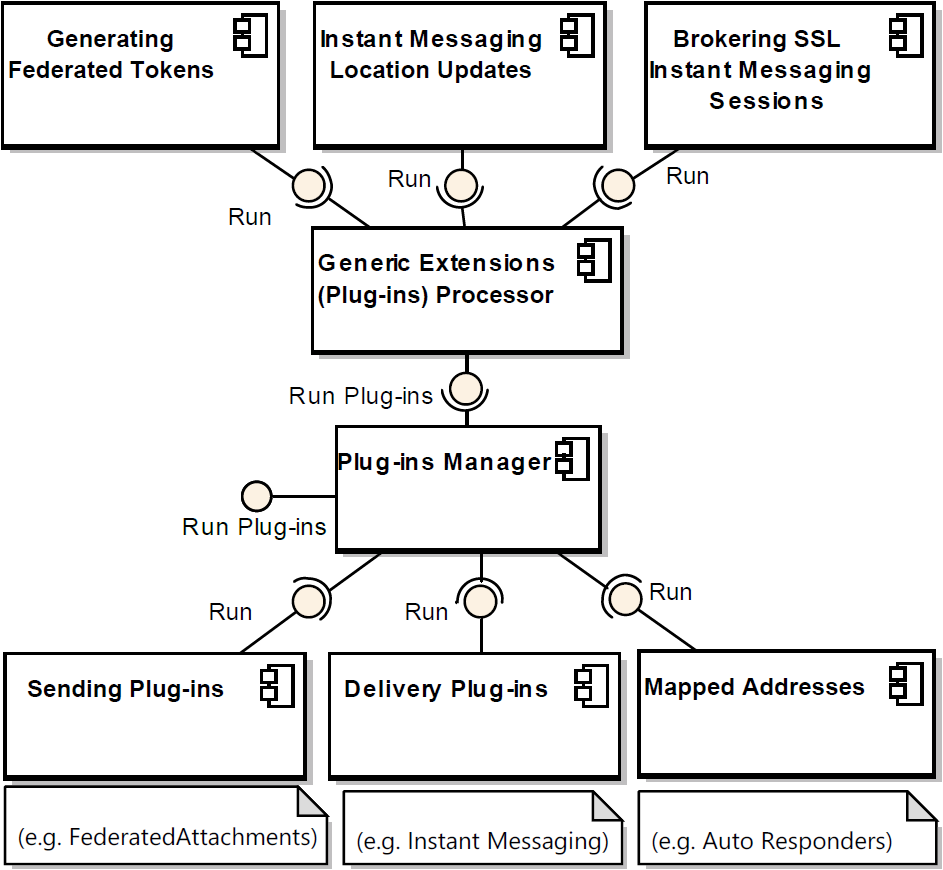}
\caption{Component diagram for some example plug-ins.}\label{fig:server-plugins-component}
\end{figure}

All server plug-ins implement the IServerPlugin interface, which allows the server to understand the plug-in's purpose and add it to the appropriate processing queues. There are two main classes of plug-ins in WSEmail: message dependent and RPC-like (or message-independent).  \tr{Some example plug-ins and their classifications are shown in Figure~\ref{fig:server-side-plugins-examples}.}{The component diagram for some example plug-ins is shown in Figure~\ref{fig:server-plugins-component}.}

\begin{figure}[ht]
\centering
\includegraphics[width=\columnwidth]{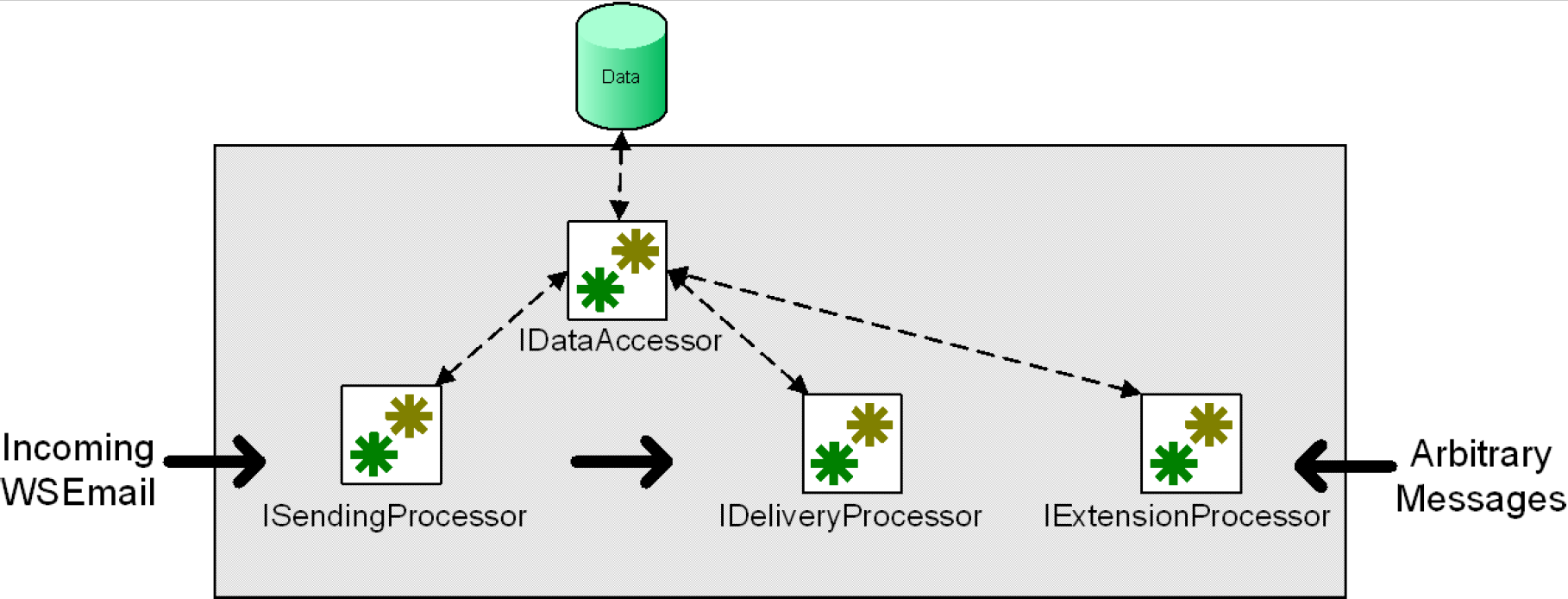}
\caption{An overview of server-side plug-ins}\label{fig:server-side-plugins}
\end{figure}

Message dependent plug-ins require that a message be present to execute. They can be inserted in various places in the delivery cycle, including at initial receipt of a message (ISendingProcessor) or at the final destination of a message (IDeliveryProcessor). ISendingProcessor plug-ins perform processing similar to that done by sendmail (\eg verifying relay permissions, stripping oversized attachments) in regular email systems. IDeliveryProcessor plug-ins act similar to user-space programs such as procmail or vacation messaging scripts in regular email systems. Figure~\ref{fig:server-side-plugins} shows the interactions and data flow of incoming WSEmail and extension requests.

RPC-like plug-ins implement a generic ``catch-all'' interface (IExtensionProcessor). On initializing, the pl\-ug-ins provide the server with an \emph{extension identifier}. It is used by the server to route incoming requests to the appropriate plug-in. There is no required or defined structure for the requests.  Most plug-ins view requests as just XML documents or fragments, providing flexibility in terms of the data an application can process.

The server plug-ins are run after the core server has performed message authentication.  The server authenticates a message by examining its attached security tokens such as X.509 certificates, username signatures, or federated identity certificates.  If they are valid, the message is entered into a queue to be processed by the appropriate plug-ins.  In the case of IExtensionProcessor and IDeliveryProcessor plug-ins, an ``environment'' object is created and passed to the plug-in to allow access to authentication tokens and raw XML streams from the server. Other plug-ins are only given the message that triggered their execution.

Plug-ins can implement more than one interface, which can increase functionality. Some plug-ins are \emph{composite}, implementing both IExtensionProcessor and IDeliveryProcessor (or any variant), allowing them to interact with messages as they are delivered, but also to be configurable using a protocol that interacts with the IExtensionProcessor interface. Implementing multiple interfaces allows plug-ins to share data that should be accessible through multiple paths. Section~\ref{sec:applications} shows examples of useful composite plug-ins.

We implemented some server plug-ins in WSEmail that would be typical for enterprise applications: data store access (IDataAccessor), database connection management (IDatabaseManager) message queues (IMail\-Queue), and local delivery (ILocalMTA).

\subsection{Client-side Plug-ins}

Client-side plug-ins affect message reading and are more dynamic than server-side plug-ins. Like server-side plug-ins, client-side plug-ins are code libraries, but for ease of implementation extend an abstract class (DynamicForms.BaseObject) instead of implementing an interface. They contain more information than server-side ones, including version and network location information. With the additional information, a plug-in can create messages that can be processed by clients that do not have the plug-in installed.  This is accomplished by having a stub executed by the receiving client download the plug-in from its supplied network location (after user approval).  Plug-in code is self-signed using Microsoft Authenticode. Plug-in information (\ie version, name, library location) is saved in a local registry to let the recipient use the plug-in again later.

\begin{figure}[tb]
\centering
\includegraphics[width=\columnwidth]{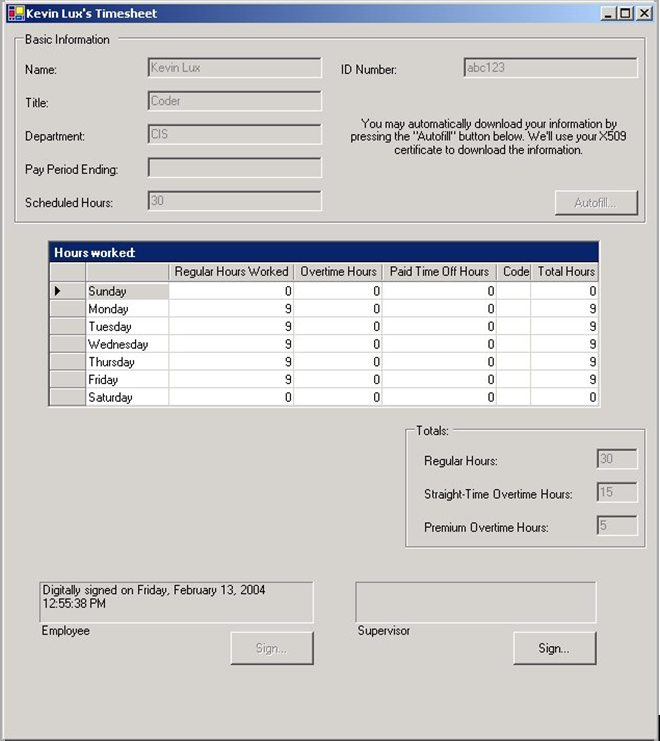}
\caption{Sample application (a timesheet)}\label{fig:timesheet}\end{figure}

A plug-in's user interface is up to the plug-in designer. A plug-in can have no interface at all, a few message boxes, or a rich GUI. Since WSEmail is based on the .NET framework, plug-in designers can use .NET's UI elements without the need to transfer additional graphics libraries. A sample plug-in interface is shown in Figure~\ref{fig:timesheet}. Plug-ins also have access to authentication information in the mail client and may petition users for access to their federated token. This allows plug-ins to perform secure web service calls to perform functions such as automatically filling in information.

\tr{For convenience and to limit bandwidth use, m}{M}ultiple plug-ins can be contained within one library file. Each plug-in will be enumerated (via .NET's reflection libraries) and registered with the client application. This allows system administrators to deploy one library with updated plug-ins instead of deploying each one separately.

\subsection{Sending and Receiving a Message}\label{sec:sending}\label{sec:receiving}

\begin{figure}[tb]
\centering
\includegraphics[width=\columnwidth]{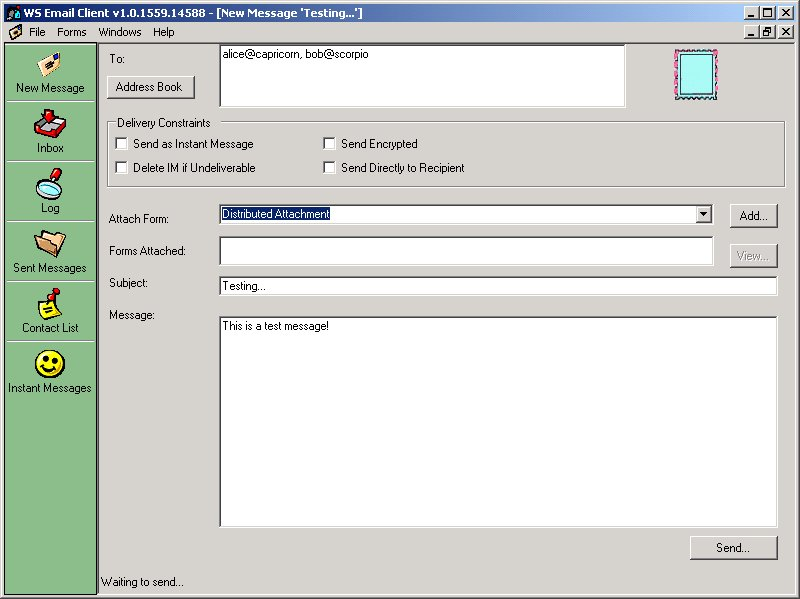}
\caption{New message screen}\label{fig:new-message}\end{figure}

\begin{figure}[tb]
  \centering
  \includegraphics[width=\columnwidth]{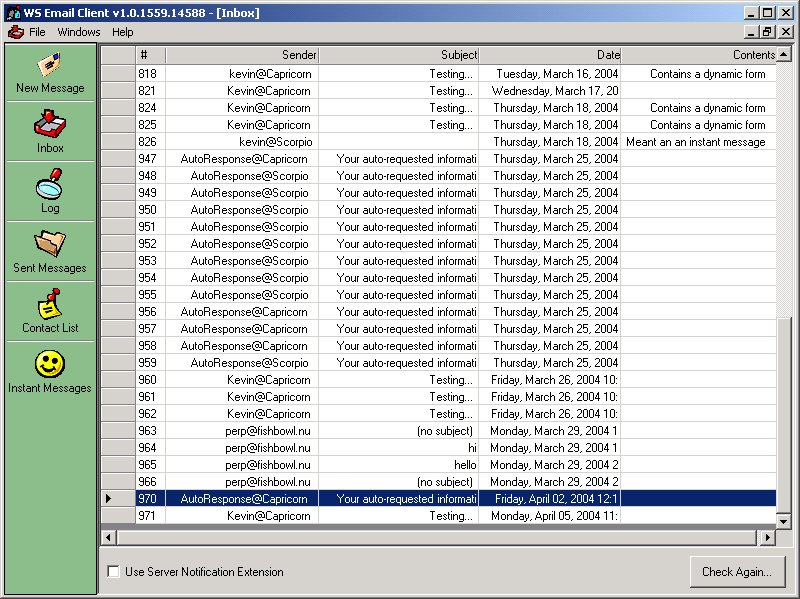}
  \caption{Inbox screen}\label{fig:inbox-screen}
\end{figure}

\begin{figure}[tb]
\centering
\includegraphics[width=\columnwidth]{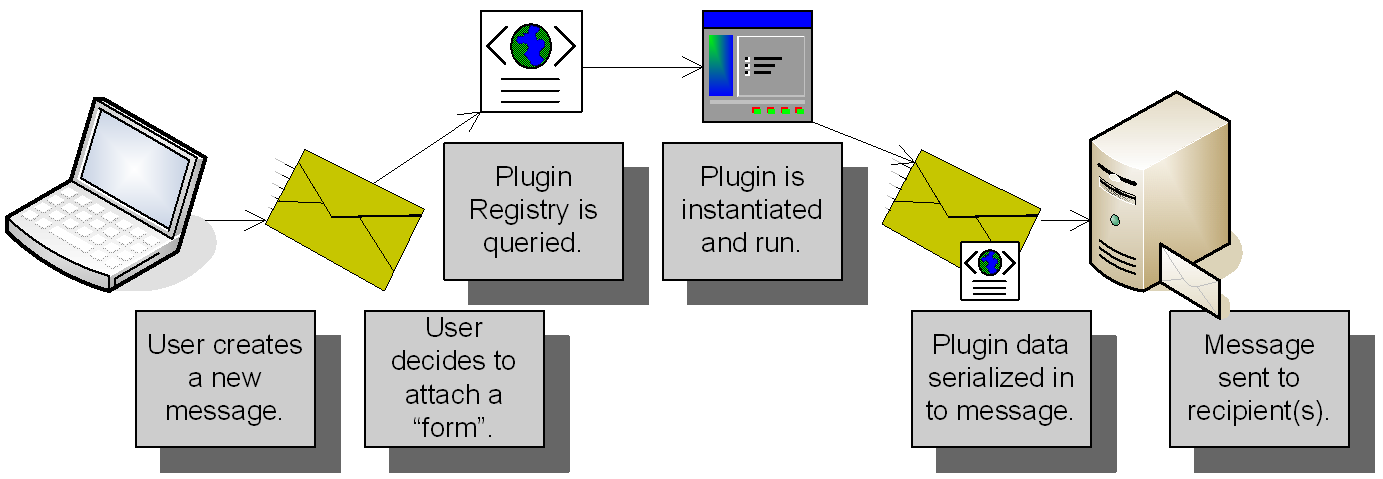}
\caption{Client-side plug-ins from the sender's perspective}\label{fig:sending-plugins}\end{figure}

When Alice wants to send a message to Bob that uses a plug-in, she attaches a ``form'' from her registry to her message (Figure~\ref{fig:new-message}). The form presents Alice with a user interface to fill in relevant information. After filling out the form, the information is serialized to an XML document and attached to the message. The original message is notified by the plug-in where it should be sent next to be appropriately processed; in this case, Bob's inbox. The plug-in is unloaded and a flag is set on the message header indicating the presence of a form. The flag is displayed in Alice's list of sent messages and Bob's inbox on receipt, allowing them to see which messages in their inbox contain forms without needing to download all of the message's contents (see Figure~\ref{fig:inbox-screen}). Figure~\ref{fig:sending-plugins} illustrates the process.

\begin{figure}[tb]
\centering
\includegraphics[width=\columnwidth]{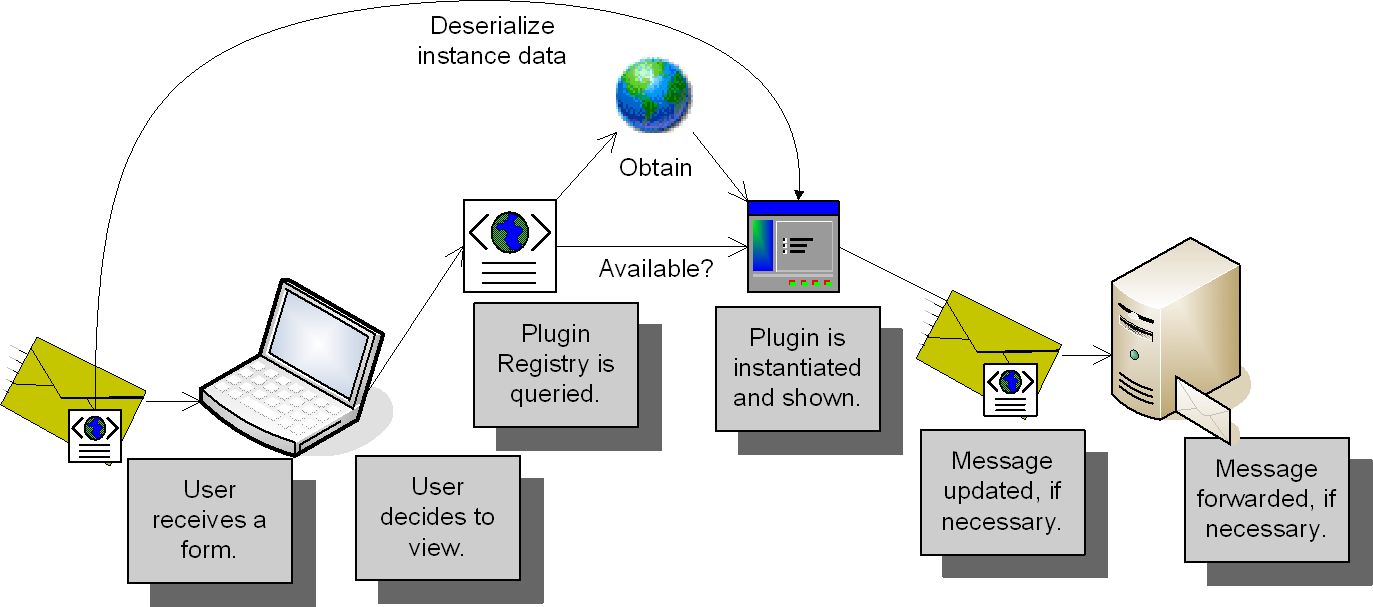}
\caption{Client-side plug-ins from the recipient's perspective}\label{fig:receiving-plugins}\end{figure}

When Bob's server receives Alice's message, he sees the new message appear in his inbox with an annotation indicating the message contains an attached form. He can view the message normally, but can also view the attached form. When Bob tries to view the form, the mail client attempts to load the appropriate plug-in or download it if it's not already present using information contained in the recipient's plug-in registry and information in the form. After the plug-in is loaded, the XML document containing the payload of the form is pushed to the plug-in that deserializes the data and loads necessary state. The plug-in then takes control, displaying a GUI that includes the information Alice sent. Figure~\ref{fig:receiving-plugins} demonstrates the typical flow for the recipient of a message with a form attached.

\section{Applications}\label{sec:applications}

WSEmail allows rich XML formats, extensible semantics on clients and routers, and a range of security tokens.  Since there are substantial development platforms for these features from software vendors, we can use WSEmail as a foundation for a suite of integrated applications that share common code, routing, security, and other features.  To illustrate, we sketch three applications that we implemented with our prototype system.  They are examples of a way forward that integrates email and IM and improves security and flexibility while avoiding the pitfalls of becoming a walled garden.

\subsection{Instant Messaging}

\begin{figure}[ht]
\centering
\includegraphics[width=\columnwidth]{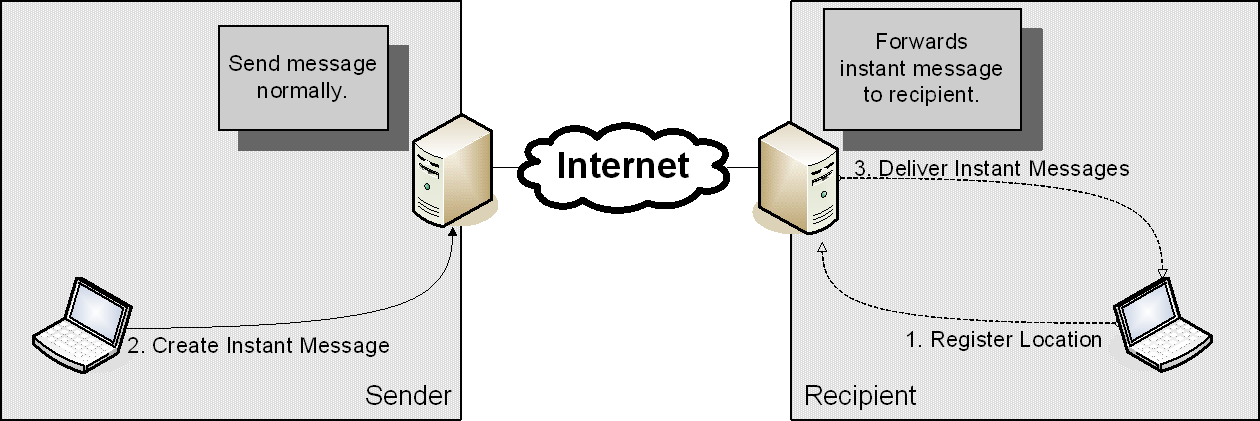}
\caption{Instant Messaging Overview}\label{fig:IM-overview}\end{figure}

\begin{figure}[tb]
  \centering
  \includegraphics[width=\columnwidth]{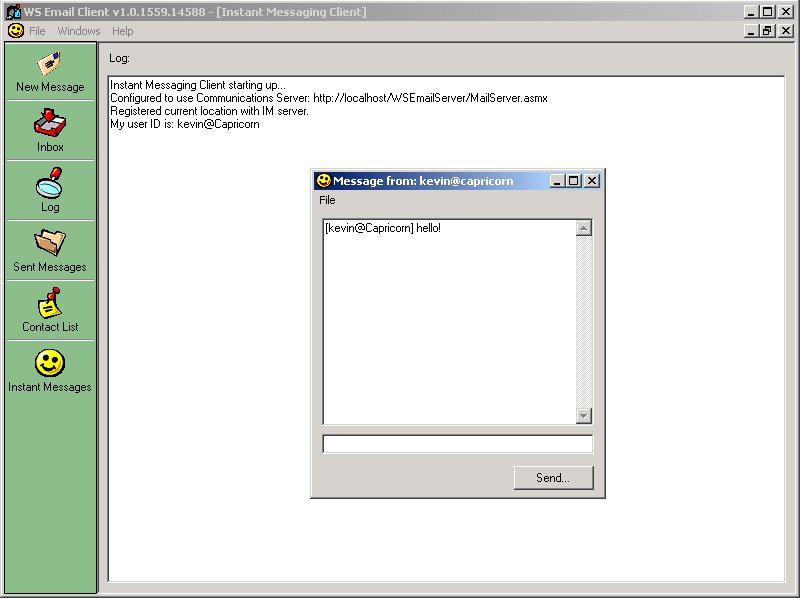}
  \caption{Instant Messaging screen}\label{fig:im-screen}
\end{figure}

IM is a messaging technique like email, but is intended for communicating short messages synchronously.  An overview of a standard IM architecture is shown in Figure~\ref{fig:IM-overview}. IM is typically disjoint from email, using different clients, servers, routing, and security.  This is unfortunate since the two messaging systems have many things in common.  We experimented with integrating the two by allowing WSEmails to be marked as instant messages (see Figure~\ref{fig:new-message}).  Such messages are posted directly to a window on the recipient client by the client server (see Figure~\ref{fig:im-screen}), subject to an access control decision.  Our implementation uses the same client, server, software, and security as the email functions.  There is an option that allows multiple parties to use TLS tunnels to a single server.

WSEmail's IM plug-in is a composite plug-in, implementing IDeliveryProcessor and IExtensionProcessor.  When the IM client program is started, the user automatically registers her location on the server using the IExtensionProcessor interface. The plug-in records the location information in an internal table. Later, when the server receives a message flagged as an instant message, it passes delivery control of the message to the instant messaging plug-in using the IDeliveryProcessor interface. The plug-in consults its table of user locations and sends the message directly to the client if it finds a match. If no match is found, the plug-in relinquishes control of the message, passing it back to the server. The server can then attempt delivery using a different matching plug-in.

Instant messages are sent to clients using Microsoft .NET Remoting (now superseded by Windows Communication Foundation). Similar to Java's RMI, Remoting allows an object to be distributed across a network. Clients remote their IM queues and have a separate thread watch it. As the server pushes messages into the queue, the client watcher thread pulls them out and displays them in a conversation-like interface.

After a conversation has been established, users can change to a synchronous channel.  An additional server-side plug-in coordinates the shift from asynchronous WSEmail messages to a ``party line'' secured with TLS. At the user's request, the plug-in allocates a TCP port running TLS and notifies all participants in the conversation of the available channel. The TLS ports do not have to be opened on the mail server, so it is possible for the mail server to act as a broker and pass the connection request on to a secure chat server farm. The recipients of the invitations are given a choice to accept the channel conversion. When connecting to the secure chat server, the clients are presented with the server's X.509 certificate and are asked to present their own certificates for authentication.  Clients who do not provide a certificate are not able to join the new secured chat session.

The secure channel instant message brokering allows users to create secure channels with arbitrary groups of people. Since the messaging flows over the TLS channel, they do not have the lag of the composition and forwarding of a WSEmail message. If the users have an existing X.509 architecture, they can easily authenticate to each other. If they lack a shared X.509 architecture, everyone (including the server) will need to set up certificate trust relationships.

\subsection{Business Workflows}

Many organizations carry out their business process workflows (\ie forms) using web forms or other web techniques.  In implementing such a system, there is a choice between a centralized system where a single web server is used by everyone and a decentralized system where information is routed by email. Email systems work best with loosely described workflows and loosely coupled participants, such as ad hoc collaborations between enterprises where neither organization can carry out all functions on a web server managed by the other party.

\begin{figure}[ht]
\centering
\includegraphics[width=\columnwidth]{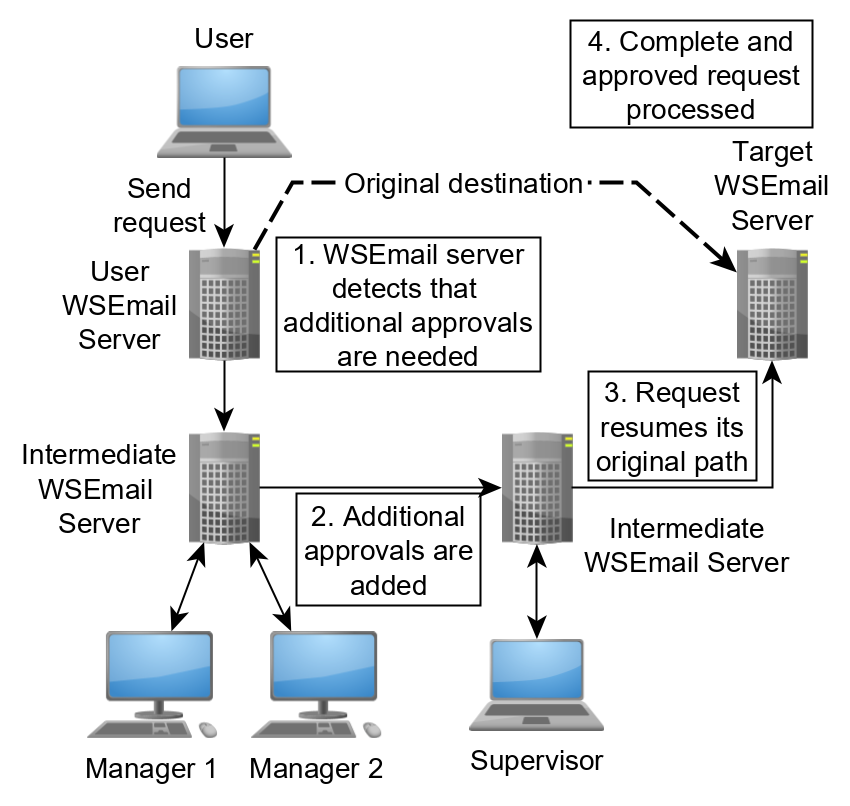}
\caption{Sample Workflow Scenario}\label{fig:sample-workflow}\end{figure}

WSEmail supports workflow management using rou\-ted forms, attachments that are sent to particular people or roles in a specified order. Routed forms use client-side plug-ins to create a rich user interface. The forms are designed to look like their paper counterparts.  We created prototypes for time sheets (Figure~\ref{fig:timesheet}) and requisition forms, including an interface to enter the required data and rules that specify which roles in the organization must ``sign off'' to approve a form. A sample workflow scenario of a form that must be passed through a chain of recipients to receive final approval is shown in Figure~\ref{fig:sample-workflow}.

The sender follows the steps described in Section~\ref{sec:sending}. In our prototype, the forms are much smarter than their paper counterparts. They can, for example, provide basic spreadsheet-like functionality or automatically populate data using the user's federated token and a secure web service query to a human resources database. To address the security of the workflow, each user has a unique X.509 certificate with the certificate's common name (CN) as the user's email address. A person in the workflow signs off on the form by attaching a digital signature to the XML.  The message thereby acquires an approval list that can be verified and audited by a third party.  The verifier can use X.509 certificates to check that the data have not been tampered with and can authenticate the approval of each member in the workflow.

As an extension of the workflow model, users can delegate their responsibilities.  Delegation is done by a user providing the name or names of people who can sign off on a form instead of him. This adds a powerful automation feature. Using server-side plug-ins, a form can be received by a program that makes decisions about delegation given the current approval list and data in the form.  A common business process that could use such a feature is a requisition form. For example, a department may be allowed to buy items under a certain fixed price, but if the total price is greater than a certain amount, additional approval by a member of the purchasing department might be required.  A requisition workflow program could easily detect this condition and expedite the purchase process by automatically forwarding the form to people who can approve the purchase or to people to whom they have delegated their responsibility.

A workflow in our system can send its result to another program. The receiving program could then validate the entire form and perform the required operations (\eg file an order, perform database manipulations).  With an increasing number of online retailers exposing their order processes as web services, it is possible to automate more business functions end-to-end within a common application such as WSEmail.

Because WSEmail also functions as a decentralized system, a workflow form can extend across multiple enterprises. All of the enterprises in the workflow would need to have an agreement to trust a common certification authority (CA) or cross-trust each other, but that is the only additional configuration that is needed. With that setup, they can use routed forms in WSEmail to create multi-enterprise processes. There are many applications that could use such a setup such as negotiating prices with a supplier or gathering approvals for press releases from interdependent corporations. Since the data are contained within the email message, the question of who hosts the data and applications for the exchange is eliminated. WSEmail sends the data wherever necessary.

\subsection{On-Demand Attachments}

Attaching files to email is a simple and convenient way to send files to a group. However, there are many problems with email attachments that WSEmail sought to improve. For instance, except for certain proprietary email systems, there is no version control system for email attachments, which usually results in many message resends so that each person gets the newest version of a file. Second, the way attachments are bundled in POP3 (IETF RFC 1939) requires users to download the entire message and attachments, even if the user only wants to read the message (a problem solved in IMAP4 (IETF RFC 3501) and critical to bandwidth and power limited devices).  A common solution is to post the attachment on a secured cloud-hosted website and to send a link in the message to it. Unfortunately, this creates an administrative and security headache since attachments are stored on third-party servers and senders must set up access control rules and authentication on an external server or for recipients who are not in their administrative domain.

\begin{figure}[ht]
\centering
\includegraphics[width=\columnwidth]{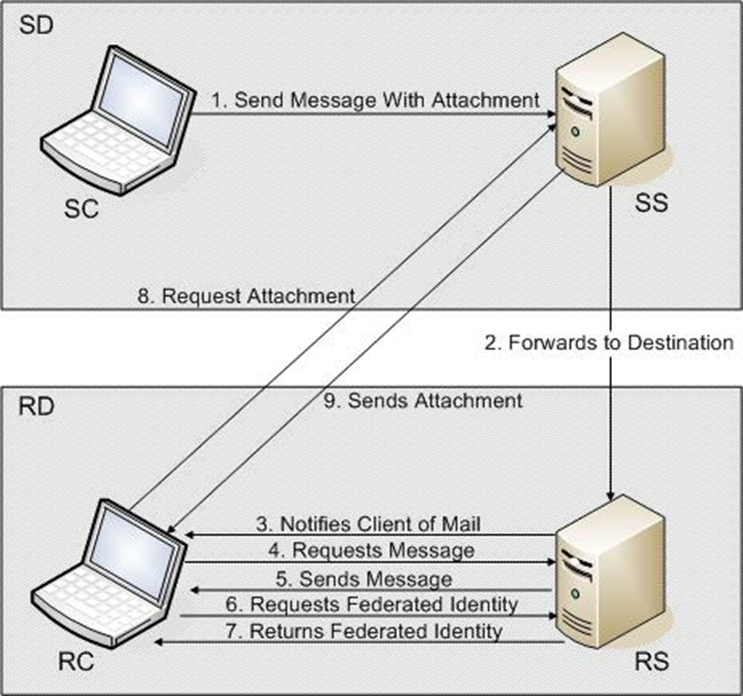}
\caption{On Demand Attachments Protocol}\label{fig:On-Demand-Attachments}\end{figure}

WSEmail solves the problem by introducing the concept of ``on-demand'' attachments. Simply, message attachments are handled as a plug-in to the WSEmail base protocols. The plug-in is composite, implementing both IExtensionProcessor and ISendingProcessor. The client creates a message that contains information about the attachment such as its size, description and a SHA1 hash in addition to the normal message fields. The request to the server contains the message as normal, along with the attachments using \tr{Direct Internet Message Encapsulation (DIME)}{a binary encoding} format. When the message is received by the server, the request is intercepted by the ``on-demand'' plug-in using the ISendingProcessor interface. The plug-in gains access to the \tr{DIME}{binary encoded} attachments in the requests. The attachments are stripped from the request and saved to a database along with a list of all the recipients. \emph{Globally Unique Identifiers (GUIDs)} are generated and injected in to the original message such that each GUID relates to one of the \tr{DIME}{binary encoded} attachments. Delivery of the message now continues normally.

A user who receives a copy of the message will see that a file is attached, but will not have a copy of it yet. The user just has the GUID for the file and the location from which it can be obtained. If the user decides to retrieve the attachment, she first acquires a federated identity token from her server. The token is presented using the IExtensionProcessor interface to the server that originally stripped the attachment along with the GUID. The server verifies the authenticity of the token and that the supplied user's token is permitted access to the GUID. If there is a match, the server sends the attachment back to the requestor as a \tr{DIME}{binary encoded} attachment.  An overview of the protocol is shown in Figure~\ref{fig:On-Demand-Attachments}.

We specified the on-demand attachments protocol formally using the TulaFale~\cite{Tulafale04} specification language, which has constructs for public key signatures and salted password authentication.  The TulaFale script compiles to a script that is verifiable with the ProVerif protocol verifier of Bruno Blanchet~(version 1.11)~\cite{Blanchet01}.  With this we were able to prove the following correspondence theorem for on-demand attachments:
\textit{if a receiving client (RC) retrieves an on-demand attachment with SC (sending client) as its return address, then SC sent the attachment}.  Details of the proof and its construction can be found in Lux~\etal~\cite{LuxMBG05} and in the online appendix.

\section{Experiments}\label{sec:experiments}

SOAP web services have been criticized for being slow due to their verbose XML formats and implementation decisions~\cite{PautassoZL08}. Security and flexibility sometimes lead to performance challenges as well. Hence a secure, flexible implementation of messaging based on web services may raise concerns about throughput and bandwidth utilization. We tested our WSEmail prototype to investigate these concerns and to illustrate the benefits of flexibility. To evaluate efficiency, we built a test bed to stress test the client and server applications and their communication protocols. In this section we describe the implementation, test bed, and experimental results.

Our code can be downloaded from \code.

We simulated an email environment where many users share an email server. They can send messages to others in the local domain or in an external domain. A user may also interact with her inbox, viewing and deleting messages. For our test, we defined four standard email operations: send, list, retrieve, and delete as we discuss below.

\subsection{Implementation}

Our WSEmail prototype runs on Windows server and client systems. Version 1.0 was implemented on the .NET framework version 1.1 and relies on the Web Services Enhancement (WSE) 1.0, CAPICOM 2.00, SQL Server 2000 (to store server messages), and IIS 5.0. The current version consists of 68 interfaces and 343 classes organized into 30 projects (see Appendix~\ref{sec:uml-class-diagrams} for UML models). Most of the software is C\# managed code created with Microsoft Visual Studio. Unmanaged code was needed to gain access to lower-level DNS functions to query for SRV records. Our IM system uses a TLS package from Mentalis (\url{mentalis.org}) since the .NET 1.1 platform does not provide native support for server TLS sockets. In 2004, we upgraded WSEmail to version 1.1 to get WS-Policy~\cite{ws-policy} support from Microsoft WSE 2.0. \tr{This was challenging because primitive functions from WSE 1.0 that we needed for WSEmail 1.0 were removed from the WSE 2.0 package forcing us to use both WSE 1.0 and WSE 2.0 for WSEmail 1.1.}{}

WSEmail uses DNS SRV records (IETF RFC 2782) to determine routing. This lets us run WSEmail over other protocols without changing the way DNS is quer\-ied.  We can also exploit the priority and weight attributes in the records. The properties of SRV records allow for future enhancement and present-day configuration that is similar to how SMTP is deployed.

\subsection{Test Bed}

\begin{figure*}[ht]
\centering
\includegraphics[width=5.7in]{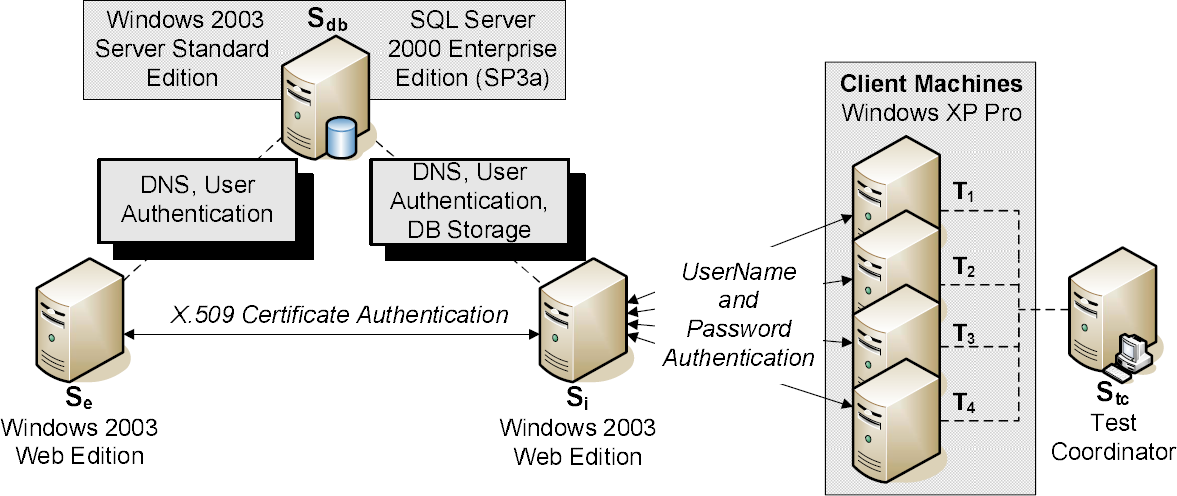}
\caption{Testbed Architecture}\label{fig:testbed}\end{figure*}

Our test bed consisted of four clients, two mail servers (designated local and external), one test coordinator, and one database/DNS server. The arrangement of the test bed is shown in Figure~\ref{fig:testbed}.

The test clients (T1--T4) performed operations by sending requests to the ``local'' email server, $S_i$. The test client actions were coordinated by the test coordinator, $S_{tc}$. There was a second server, $S_{e}$, that acted as both the ``external'' email server and a load generator for the ``local'' system. $S_{db}$ hosted a message storage database and DNS records for $S_{i}$ and $S_{e}$. The clients all had Pentium 4 2.8GHz processors with 512MB of memory and Windows XP Professional. They performed four operations during the test execution: (1) send a message to a recipient, (2) list the headers of messages in the client's inbox, (3) retrieve a message, (4) delete a message.

The test coordinator, $S_{tc}$, was responsible for distributing the test specifications, starting the test, and receiving results from each client. $S_{tc}$ broadcasted its network address, instructing all clients to connect to it and download the test specifications file. The clients then waited for $S_{tc}$ to announce the start of the test, after which the clients executed requests to $S_{i}$ in compliance with the specifications. After each client finished, the latencies for each request were reported to $S_{tc}$.

The test specifications document described exactly what each client should do. It specified whether the client should authenticate using a username token (user name and password) or X.509 certificate. It also specified how many messages to send from the client, to whom they were to be sent, and the size of the message body. The specification document also indicated the total number of requests to send and the ratios of the four types of requests.

The local server $S_{i}$ was the focus of our test. It accepted incoming messages from the clients and $S_{e}$. It performed the necessary authentication and forwarded external messages to the appropriate destination after performing DNS resolution. If the destination was local (\ie the recipient was on $S_i$), the message was stored in $S_{db}$. If the destination was external, the message was forwarded to $S_{e}$. We allowed the local and external server to share a database and DNS server since they were not performance bottlenecks in the system.

$S_{e}$  played two roles in our test bed. First, it imitated the entire external client list, so all emails directed to any external client were forwarded to it. On receiving a message addressed to one of the clients that it simulated, it did not save it to the database server. We did this to prevent $S_{i}$ from experiencing extra latency due to $S_{e}$'s database transaction. Rather, it performed the required certificate checking to verify authenticity and then discarded the message. Second, it acted as a load generator and sent one message per second addressed to each of the four clients, T1--T4. The messages were all received by $S_i$, authenticated, and stored in $S_{db}$.

\subsection{Procedure and Results}

The test coordinator $S_{tc}$ provided a test specification document that instructed each client to run one execution thread sending 2,000 requests to $S_i$. The clients chose send, list, retrieve, and delete operations with 25\% chance. In cases where the delete operation was to be performed on an unpopulated inbox, it was considered a no-op and not counted towards the results. To avoid this condition, each client's inbox was primed with about six messages. To get the most out of each send event, each message was addressed to both a randomly chosen local client and an external client. The clients were all instructed to authenticate to $S_{i}$ using username token authentication. $S_{i}$ and $S_{e}$ authenticated to each other using X.509 certificate signing. The duration of the test was 1830 seconds.

To get a client-side view of the efficiency of the system, we measured the latency of each request. A timer was started as the client contacted $S_{i}$ with a request and stopped after the client received the appropriate response (\eg inbox listing, message received confirmation). The time difference between the client's request and the server's complete response was the latency of the operation. The results of this calculation were an average of 284 ms per request with a variance of 138.9 ms. The minimum and the maximum latencies were 46.876 ms and 4,000 ms (4 s) respectively. Note that the message received confirmation does not mean that the message was delivered to the final recipient, just that the message was placed in the delivery queue.

The test results in Table~\ref{tab:client-si} show the data volume (in MB) sent between the clients and $S_i$ broken down by email operation. The total data sent from the clients to $S_{i}$ was 25.98 MB and from $S_{i}$ to all clients 155.22 MB.

\tr{
\begin{table*}[tb]
\centering
\begin{tabular}{|l||l|l|l|l|}\hline
Operation		             & Send	& List	& Retrieve	& Delete\\\hline\hline
\# of requests	             & 1970	& 2024	& 2026	    & 1980\\\hline
\% of all requests           & 24.6	& 25.3	& 25.4	    & 24.7\\\hline
Client$\rightarrow$Server Data (MB)    & 10.74& 8.42	& 8.62	    & 8.4\\\hline
Server$\rightarrow$Client Data (MB)	 & 12.31& 324.55& 20.41	    & 12.08\\\hline
\end{tabular}
\caption{Data volume sent between clients and $S_i$}\label{tab:client-si}\end{table*}}
{
\begin{table}[tb]
\centering
\caption{Data volume sent between clients and $S_i$}\label{tab:client-si}
\begin{tabular}{lllll}\toprule
Operation		             & Send	& List	& Retrieve	& Delete\\\midrule
\# of requests	             & 1969	& 2014	& 2037	    & 1980\\\midrule
\% of all requests           & 24.6	& 25.2	& 25.4	    & 24.7\\\midrule
Client$\rightarrow$Server    &      &       &   	    & \\
Data (MB)                    & 8.31 & 5.83	& 5.99	    & 5.85\\\midrule
Server$\rightarrow$Client 	 &      &       &           & \\
Data (MB)	                 & 4.06 & 127   & 20.33	    & 3.83\\\bottomrule
\end{tabular}
\end{table}
}

Since each message was also sent to an external client, each send action also sent a message from $S_{i}$ to $S_{e}$. The data sent from $S_i$ to $S_e$ are summarized in Table~\ref{tab:si-se}. The total data sent from $S_{i}$ to $S_{e}$ was 33.74 MB and from $S_{e}$  to $S_i$ was 31.55 MB.

\tr{
\begin{table*}[tb]
\centering
\begin{tabular}{|l||l|l|l|l||}\hline
Server Name  & \# messages & Sent (MB)	& Received Confirmations (MB)\\\hline
$S_i$		        & 1970	        & 2024	    & 2026\\\hline
$S_{e}$ 		        & 24.6	        & 25.3	    & 25.4\\\hline
\end{tabular}
\caption{Bytes sent between $S_{i}$ and $S_{e}$ }\label{tab:si-se}\end{table*}
}{
\begin{table}[tb]
\centering
\caption{Data volume sent between $S_{i}$ and $S_{e}$ }\label{tab:si-se}
\begin{tabular}{lllll}\toprule
Server & \# & Sent	& Received \\
Name   & messages & (MB)	 & Confirmations (MB)\\\midrule
$S_i$  & 1969	 & 20.63 & 13.11\\\midrule
$S_{e}$& 1830    & 19.38 & 12.17\\\bottomrule
\end{tabular}
\end{table}
}

The test bed data transfer was recorded using server-side network monitor tools run at $S_{i}$ and $S_{e}$. The TCP sessions were reconstructed using tcpflow (\url{https://github.com/simsong/tcpflow}) and processed with a\-wk and Perl. Since $S_{e}$, acting as a load generator, sent one message per second, 1830 messages were sent from $S_{e}$ to $S_{i}$ during the test. The corresponding byte count represents the messages that were sent and the notification messages that were received.

\subsection{Analysis}

A best-case test of SMTP with no load on the server or network and no contention for resources yielded an average latency of 170 ms to send a message of about the same size as the WSEmail messages we sent in our experiment. The average difference in latency between WSEmail and the SMTP test is 114 ms, which accounts for the additional overhead of the XML parsing and cryptography. In that time span a large number of operations took place: one secret key signature, one private key signature verification, two public key signatures, and one public key signature verification. Since the entire system uses XML, we conclude that performance is not a barrier to secure web services in this type of application. The extra latency would likely be unnoticeable a typical client/server environment.

XML and XMLDSIG have drawbacks due to their verbosity. Our test bed sent 1 KB mail messages that ballooned to 10 KB responses to the retrieve message action to make XMLDSIG work. At least 30\% of those bytes were the Base64 encoded representations of the certificates used for signing messages. After the certificate size, the WS-Security~\cite{ws-security} structures were a significant amount of overhead, accounting for about 30\% of the bytes. The overhead could be reduced by using an alternative certificate distribution mechanism or via techniques to reduce SOAP's overhead~\cite{Senagi2016}. \tr{It might also be useful to examine how messages are signed to minimize their verbosity.}{}

Our experiments bode well for web service efficiency, especially for high volume messaging. Extending our experimental results, WSEmail is theoretically capable of handling approximately 1787 messages a minute (combination of incoming and outgoing). We looked for published benchmarks to compare this against and found that the University of Wisconsin-Parkside 
had a peak usage of 1716 (total of incoming and outgoing) messages per minute in 2005, meaning it should be possible for a single WSEmail server similar to our test system to handle the normal load at that institution. 



\section{Related Work and Similar Systems}\label{sec:related}

Work related to WSEmail can be divided into three general areas: email improvements, analysis of web service security, and improved rich internet messaging systems.

\paragraph{Email Improvements} Improvements to the SMTP messaging system have been motivated by two, sometimes overlapping goals: strong message authentication and spam prevention. PGP offers authentication tools that include public/private key signing and encryption. Privacy Enhanced Mail (PEM) (IETF RFCs 1421--4) has mechanisms for privacy, integrity, source authentication, and nonrepudiation using public and private key encryption and end-to-end encryption. Zhou~\etal~\cite{ZhouKOC99} use formal tools to verify the properties of PEM.  Abadi \etal~\cite{AbadiGHP02} use a trusted third party to achieve message and source authentication and formally prove correctness of their messaging protocol.

Changes to the SMTP system aimed at spam reduction include DomainKeys Identified Mail (DKIM) (IETF RFC 6376) that uses public key cryptography and an option for server signed (rather than client-signed) messages and Petmail~\cite{petmail05}(~\url{http://petmail.lothar.com/}). Petmail uses the GPG encryption utility for public key encryption and signing of messages. Users are identified by IDRecords, self-signed binary blobs that include public key, identity, and message routing information. Petmail agents can enforce IDRecord whitelists and policies for contact from first time senders. First-time senders may be forced to obtain tickets from a third-party Ticket Server that may check to ensure that the sender is a human (using CAPTCHA reverse Turing tests). Messages can be encapsulated and sent using SMTP, Jabber, or another queuing transport protocol. Patterns and options for sender anonymity are offered as well. Our most recent work on extensions of WSEmail show how to match some of Petmail's capabilities and perform others not supported by Petmail using WS-Policy~\cite{ws-policy} negotiations and dynamic plug-ins.


\paragraph{Web Service Security} WS-Security~\cite{ws-security} is a set of protocols for adding security enhancements to SOAP messages. Regarding web services security analyses, the Samoa project at Microsoft Research 
developed a formal semantics for proving web services authentication theorems~\cite{BhargavanFG04} and the TulaFale language for automating web service security protocol proofs~\cite{Tulafale04}.  Based on their ideas and others, we performed follow-up work based on WSEmail, including adaptive middleware messaging policy systems~\cite{AfandiZG06}, attribute-based messaging~\cite{BobbaFKGK06}, and a secure alert messaging protocol~\cite{GioachinSMGS07}.

\subsection{Rich Messaging Systems}

Several vendor-specific all-in-one internet messaging systems have been developed recently, including Google Talk, Skype, WeChat, Slack, and WhatsApp.  All provide security at the expense of openness.  Some allow extensions and integrated apps, but only via a centralized service.

\paragraph{Gmail and Hangouts}  Google's Gmail platform evolved from an email system to include a chatting service called Google Talk in 2005 and a unified platform called Hangouts in 2013.  The talk application has similarities to WSEmail in that it integrates IM with email messages (allowing conversion between the two) and integrates with other chat protocols such as Jabber.  Talk differs from WSEmail in that it uses hop-by-hop encryption instead of end-to-end~\cite{Gtalk-secure}.  It is also primarily client-server based (via Google), although it will route calls in a peer-to-peer manner if possible~\cite{HangoutsP2P}.  Talk and Hangouts use proprietary communication protocols, so the platforms are not amenable to third-party extensions.

\paragraph{Skype} Skype integrates voice and chat into one app.  Chat messages sent to offline users are sent like emails, stored on the server and delivered to the target at next login.  Skype's security model is similar to WSEmail's secure chat architecture in that it uses public key encrypted messages and challenges for authentication, establishes a shared key, and then uses the shared key to create a secure end-to-end channel.  It uses proprietary protocols and does not offer integration with other chat or voice communication tools~\cite{SkypeSecurity}.

\paragraph{WeChat} WeChat is a service that provides instant messaging and chat.  Its communication protocol is proprietary, but forensic analyses and protocol analyses have found that its communication protocols are server based and encrypted using a custom combination of a fixed RSA key and derived AES keys~\cite{HuangLHQH15,Paleari13}.  WeChat allows integration of miniprograms via its central servers.

\paragraph{Slack} Slack is a cloud based tool for team collaboration that offers chat and document storage.  Its security protocols are based on TLS 1.2, SHA2, and AES and support enterprise specific keys~\cite{SlackSecurity19}.  Protocol details are proprietary, but black box testing and protocol analysis have shown them to be server based with no peer-to-peer or direct connections~\cite{Iwase16}.  In contrast to others, Slack allows \emph{bots}, software agents that listen to conversations and data and act based on them.

\paragraph{OTR}  The Off the Record (OTR)~\cite{BorisovGB04} protocol introduces a mechanism for secret, authenticated low-latency communication that preserves the ability for participants to repudiate their messages later.  OTR has been installed in a number of operating systems and secure chat tools such as cryptocat~\cite{KobeissiB13}.

\paragraph{WhatsApp} WhatsApp is a cloud based chat and video call service aimed at smartphones.  The WhatsApp client authentication protocol underwent significant changes between versions.  The pre-2016 WhatsApp client authentication steps are as follows~\cite{Anglano14,KarpisekBB15}.  At installation time, a shared password $pw$ is generated for the user account and stored on the mobile device \tr{in \url{/data/data/com.whatsapp/files/pw} on the device and }{before being} transferred to the WhatsApp servers.  \tr{Login then consists of the following steps:

\begin{enumerate}
  \item The client sends an initial \emph{auth} message with its client number and the method it wants to use for authentication.  The message isn't encrypted.  The message includes information about the client software version and capabilities.
  \item The server responds with some parameters that include a nonce $n$.
  \item The client and server use $pw$ and $n$ to generate four keys using PBKDF2: $k_{se}$ (server encryption), $k_{si}$ (server integrity), $k_{ce}$ (client encryption), and $k_{ci}$ (client integrity).
\end{enumerate}}{It is used along with a nonce to generate session keys using a key derivation algorithm at login time.}
Since 2016, WhatsApp uses the Signal protocol, an end-to-end encryption scheme based on an elliptical curve public/private key pair generated at install time~\cite{WhatsApp16}.  It uses the Signal ratcheting protocol for instant messaging and voice communication~\cite{Signal16}.  WhatsApp is a closed source system that does not enable extensions or integration with third-party clients.

\section{Conclusion}\label{sec:conclusion}

\tr{We have explored WSEmail, the development of email functions as a family of web services, by developing a prototype system based on an architecture that emphasizes flexibility, security, and integration. We have shown that WSEmail is amenable to the addition of new protocols and the formal analysis of these protocols. We have also shown that the basic WSEmail functions have satisfactory performance. In ongoing work, we are exploring several directions such as: new applications that exploit improved integration between web-like data retrieval functions and the messaging system; challenges to interoperability with a Java implementation of the MUA; and ways to express and negotiate messaging policies. For widespread use, WSEmail faces substantial problems with standardization and interoperability with SMTP, which may be mitigated by writing more plug-ins like our SMTP-compatible relay agent. However, it is well-suited to some high-security applications even now, offers ideas in exploring the general design space for Internet messaging, and can rely on the standardization advantages of XML as an aid to addressing interoperability challenges. We also aim to support WSEmail on diverse platforms. A project of Heo, Patel, and Shah was partially successful in doing this for a Java WSEmail client based on Sun's JWSDP 1.4 with X.509 security. 
}
{
Email and IM have advanced significantly in the past decade.  Email's protocols have remained stable, while server side technologies have reduced misuse (\eg spam).  IM has improved in terms of security and privacy, offering stronger authentication that increases trust.  Modern secure messaging systems, however, have achieved their gains, at the price of openness and ability to interact with each other and with extensions.  As WSEmail demonstrates, another evolutionary path was possible, one that would have offered increased security and privacy while retaining openness and extensibility.

We have detailed WSEmail, including its software architecture, communication protocols, and dynamic extensions mechanism.  It allows for rich internet messaging with flexible policy options for routing and delivery. Its messaging protocols are amenable to formal correctness and security verification.  Our performance measures show reasonable communication latency on legacy hardware and operating systems.

WSEmail serves as model for how it is possible to unify communication tools such as email and IM in a secure and private way without hindering extensibility and openness.  Next generation messaging systems would do well to take lessons from it.
}

\paragraph{Acknowledgements}  We are grateful for discussions of WSEmail that we had with Martin Abadi, Raja Afandi, Noam Arzt, Karthikeyan Bhargavan, Luca Cardelli, Dan Fay, Eric Freudenthal, Cedric Fournet, Andy Gordon, Ari Hershl Gordon-Schlosberg, Munawar Hafiz, Jin Heo, Himanshu Khurana, Ralph Johnson, Bjorn Knutsson, Jay Patel, Neelay Shah, Kaijun Tan, and Jianqing Zhang. We are also grateful to Nayan L. Bhattad for technical support with experiments.

\bibliographystyle{plain}
\bibliography{wsemail}

\begin{thebibliography}{10}

\bibitem{AbadiGHP02}
M.~Abadi, N.~Glew, B.~Horne, and B.~Pinkas.
\newblock Certified email with a light on-line trusted third party: design and
  implementation.
\newblock In {\em Proceedings of the Eleventh International Conference on World
  Wide Web}, pages 387--395. ACM Press, 2002.

\bibitem{AfandiZG06}
Raja Afandi, Jianqing Zhang, and Carl~A. Gunter.
\newblock {AMPol-Q}: Adaptive middleware policy to support {QoS}.
\newblock In {\em Proceedings of the 4th International Conference on
  Service-Oriented Computing}, ICSOC'06, pages 165--178, Berlin, Heidelberg,
  2006. Springer-Verlag.

\bibitem{Anglano14}
Cosimo Anglano.
\newblock Forensic analysis of {WhatsApp} messenger on {A}ndroid smartphones.
\newblock {\em Digital Investigation}, 11(3):201--213, 2014.
\newblock {S}pecial Issue: Embedded Forensics.

\bibitem{ws-policy}
Siddharth Bajaj, Don Box, Dave Chappell, Francisco Curbera, Glen Daniels,
  Phillip Hallam-Baker, Maryann Hondo, Chris Kaler, Dave Langworthy, Anthony
  Nadalin, Nataraj Nagaratnam, Hemma Prafullchandra, Claus von Riegen, Daniel
  Roth, Jeffrey Schlimmer, Chris Sharp, John Shewchuk, Asir Vedamuthu, \"{U}mit
  Yal\c{c}inalp, and David Orchard.
\newblock Web {S}ervices {P}olicy 1.2 - {F}ramework {(WS-Policy)}.
\newblock {W3C} member submission, World Wide Web Consortium, 25 April 2006.

\bibitem{xmldsig}
M.~Bartel, J.~Boyer, B.~Fox, B.~LaMacchia, and E.~Simon.
\newblock {XML}-signature syntax and processing.
\newblock {W3C} recommendation, World Wide Web Consortium, Feb 2002.

\bibitem{BhargavanFG04}
K.~Bhargavan, C.~Fournet, and A.~Gordon.
\newblock A semantics for web services authentication.
\newblock In {\em Proceedings of the 31st ACM SIGPLAN-SIGACT Symposium on
  Principles of Programming Languages}, pages 198--209, New York, NY, 2004. ACM
  Press.

\bibitem{Tulafale04}
K.~Bhargavan, C.~Fournet, A.~D. Gordon, and R.~Pucella.
\newblock {TulaFale}: A security tool for web services.
\newblock In {\em International Symposium on Formal Methods for Components and
  Objects (FMCO'03)}, LNCS. Springer, 2003.

\bibitem{SkypeSecurity}
Philippe Biondi and Fabrice Desclaux.
\newblock Silver needle in the {S}kype.
\newblock In {\em BlackHat Europe 2006}. Black Hat, March 2006.

\bibitem{Blanchet01}
B.~Blanchet.
\newblock An efficient cryptographic protocol verifier based on {P}rolog rules.
\newblock In {\em Proceedings of the 14th IEEE Workshop on Computer Security
  Foundations}, page~82. IEEE Computer Society, 2001.

\bibitem{BobbaFKGK06}
R.~Bobba, O.~Fatemieh, F.~Khan, C.~A. Gunter, and H.~Khurana.
\newblock Using attribute-based access control to enable attribute-based
  messaging.
\newblock In {\em 2006 22nd Annual Computer Security Applications Conference
  (ACSAC'06)}, pages 403--413, Dec 2006.

\bibitem{BorisovGB04}
Nikita Borisov, Ian Goldberg, and Eric Brewer.
\newblock Off-the-record communication, or, why not to use {PGP}.
\newblock In {\em Proceedings of the 2004 ACM {W}orkshop on {P}rivacy in the
  {E}lectronic {S}ociety}, pages 77--84, New York, NY, USA, 2004. ACM.

\bibitem{Signal16}
Katriel Cohn-Gordon, Cas Cremers, Benjamin Dowling, Luke Garratt, and Douglas
  Stebila.
\newblock A formal security analysis of the {Signal} messaging protocol.
\newblock Cryptology ePrint Archive, Report 2016/1013, 2016.
\newblock eprint.iacr.org/2016/1013.

\bibitem{GioachinSMGS07}
F.~Gioachin, R.~Shankesi, M.~J. May, C.~A. Gunter, and W.~Shin.
\newblock Emergency alerts as {RSS} feeds with interdomain authorization.
\newblock In {\em Second International Conference on Internet Monitoring and
  Protection (ICIMP 2007)}, pages 13--13, July 2007.

\bibitem{HangoutsP2P}
Google.
\newblock Peer-to-peer calling in {Hangouts}.
\newblock Hangouts Help [Online], 2018.
\newblock last accessed 22 Nov 2018.
  \url{https://support.google.com/hangouts/answer/6334301?hl=en}.

\bibitem{Gtalk-secure}
GoogleTalkGuide.
\newblock Can my {GTalk} discussion be tracked?
\newblock Google Talk Help Discussion Archive [Online], 21 Nov 2006.

\bibitem{HuangLHQH15}
Q.~Huang, P.~P.~C. Lee, C.~He, J.~Qian, and C.~He.
\newblock Fine-grained dissection of {WeChat} in cellular networks.
\newblock In {\em 2015 IEEE 23rd International Symposium on Quality of Service
  (IWQoS)}, pages 309--318, June 2015.

\bibitem{Iwase16}
Yoshimasa Iwase.
\newblock Is {Slack's WebRTC} really slacking?
\newblock webrtcH4cKS, 24 Mar 2016.

\bibitem{KarpisekBB15}
F.~Karpisek, I.~Baggili, and F.~Breitinger.
\newblock {WhatsApp} network forensics: Decrypting and understanding the
  {WhatsApp} call signaling messages.
\newblock {\em Digital Investigation}, 15:110--118, 2015.
\newblock {S}pecial Issue: Big Data and Intelligent Data Analysis.

\bibitem{KobeissiB13}
Nadim Kobeissi and Arlo Breault.
\newblock Cryptocat: Adopting accessibility and ease of use as security
  properties.
\newblock {\em CoRR}, abs/1306.5156, 2013.

\bibitem{LuxMBG05}
Kevin~D. Lux, Michael~J. May, Nayan~L. Bhattad, and Carl~A. Gunter.
\newblock Wsemail: Secure internet messaging based on web services.
\newblock In {\em 2005 IEEE International Conference on Web Services (ICWS)},
  Orlando, FL, USA, July 2005. IEEE.

\bibitem{ws-security}
Anthony Nadalin, Chris Kaler, Ronald Monzillo, and Phillip Hallam-Baker.
\newblock Web services security: {SOAP} message security 1.1 ({WS-S}ecurity
  2004).
\newblock Standard wss-v1.1-spec-os-SOAPMessageSecurity, OASIS, Feb 2006.

\bibitem{Paleari13}
Roberto Paleari.
\newblock A look at {WeChat} security, 17 Sept 2013.

\bibitem{PautassoZL08}
Cesare Pautasso, Olaf Zimmermann, and Frank Leymann.
\newblock Restful web services vs. "big"' web services: Making the right
  architectural decision.
\newblock In {\em Proceedings of the 17th International Conference on World
  Wide Web}, WWW '08, pages 805--814, New York, NY, USA, 2008. ACM.

\bibitem{Senagi2016}
Kennedy~Mutange Senagi, George Okeyo, Wilson Cheruiyot, and Michael Kimwele.
\newblock An aggregated technique for optimization of {SOAP} performance in
  communication in web services.
\newblock {\em Service Oriented Computing and Applications}, 10(3):273--278,
  Sep 2016.

\bibitem{SlackSecurity19}
Slack.
\newblock {\em Security White Paper: {Slack's} Approach to Security}, 2019.

\bibitem{petmail05}
Brian Warner.
\newblock {\em Petmail Design}.
\newblock Petmail, 20 July 2005.
\newblock last accessed 13 Aug 2019.

\bibitem{WhatsApp16}
WhatsApp.
\newblock {WhatsApp} encryption overview.
\newblock Technical white paper, WhatsApp, 16 Nov 2016.

\bibitem{ZhouKOC99}
Dan Zhou, J.~C. Kuo, S.~Older, and S.~K. Chin.
\newblock Formal development of secure email.
\newblock In {\em Proceedings of the 32nd Annual Hawaii International
  Conference on Systems Sciences}, Jan 1999.

\end{thebibliography}

\appendix





\section{UML Class Diagrams}\label{sec:uml-class-diagrams}

\begin{figure*}
  \centering
  \includegraphics[width=6.4in]{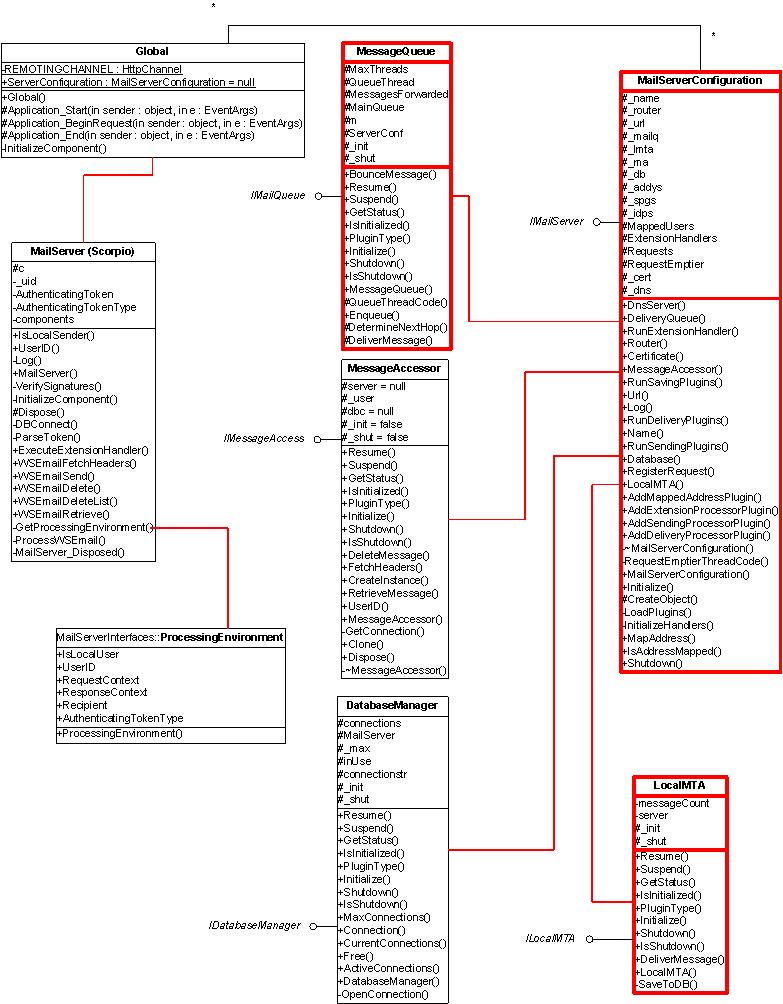}
  \caption{Server implementation class diagram}\label{fig:server-implementation}
\end{figure*}

\begin{figure*}[tb]
  \centering
  \includegraphics[width=6.25in]{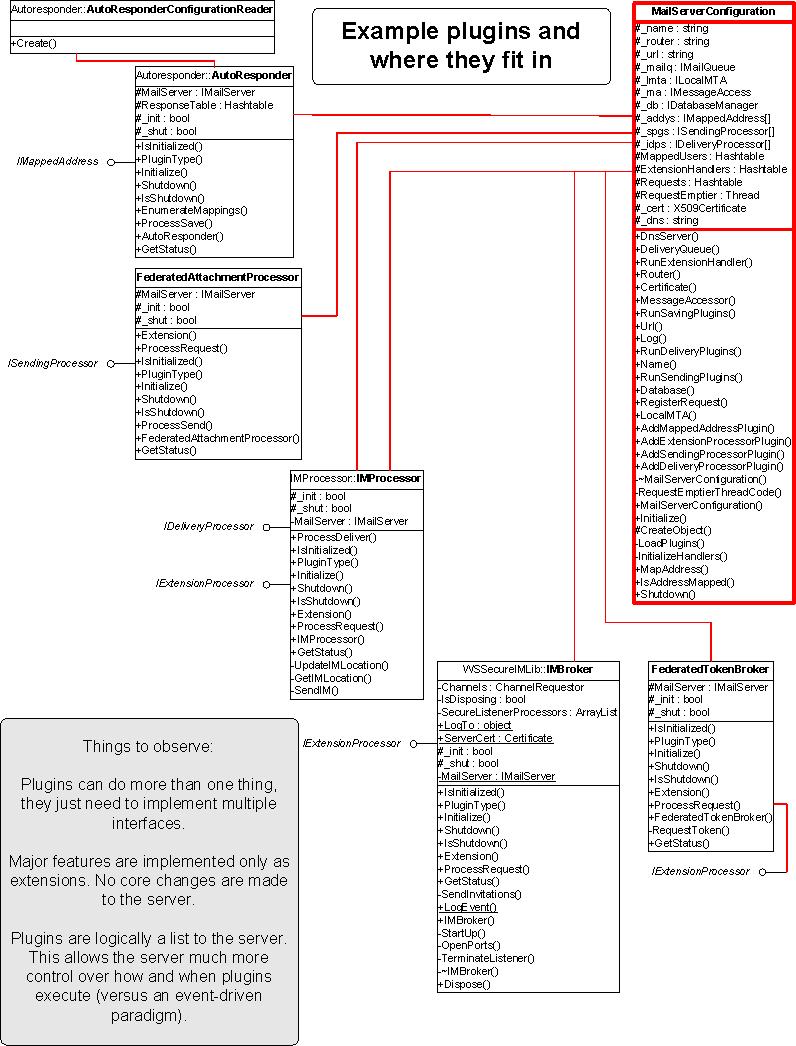}
  \caption{Server architecture with plugins}\label{fig:server-arch-plugins}
\end{figure*}

Figures~\ref{fig:server-implementation} and~\ref{fig:server-arch-plugins} show the UML class diagram for the server-side implementation and plugin architecture respectively.

\end{document}